\documentclass[aps,preprint,groupedaddress]{revtex4-1}
\usepackage{amsmath}
\usepackage{graphicx}
\usepackage{color}
\usepackage{ulem}
\graphicspath{{Figures/}}
\DeclareMathOperator{\sign}{sign}
\begin{document}
\title{The Coalescence of Leidenfrost Drops:\\Numerical Simulations using the Dynamic van der Waals Theory}

\author{M. T. \surname{Taylor}}
\email[]{Corresponding author: mtaylor@connect.ust.hk}

    \affiliation{Department of Mathematics, Hong Kong University of Science and Technology, Clear Water Bay, Kowloon, Hong Kong}
\date{\today}

\begin{abstract}
%
%
The Leidenfrost effect describes liquid drops under gravity levitating on a vapour cushion, which is sourced at the liquid-vapour interface from evaporation
caused by the hot substrate below.
It has been experimentally observed that when two or more Leidenfrost drops approach one another, there is a short-range repulsion that can prevent coalescence.
Meanwhile, non-wetting drops display an interesting property whereby they can remove themselves from a substrate spontaneously at the point of coalescence.
%
%
Since small Leidenfrost drops appear similar to non-wetting drops, the former have been used in coalescence experiments to represent the latter.
%
%
%
In the present work, we investigate the coalescence process of initially momentum-less Leidenfrost drops. This is
carried out by numerically simulating two-dimensional systems of liquid and vapour
using the recently developed dynamic van der Waals theory, in which
the two-phase hydrodynamics is coupled with liquid-vapour transition.
%
%
The cause of the increase in vertical momentum upon coalescence is found to be similar to the non-wetting case; explained by the common mechanism of a reduction in total surface energy. 
It is confirmed that Leidenfrost drops experience a remarkably strong short-range repulsion, due to a build-up of vapour in the common space under both drops. This resistive force scales in a similar manner to the pressure-sourced force from the vapour layer under a single drop, in concurrence with intuition, with good agreement seen in the simulation results. 
A threshold drop size therefore exists in order for an applied horizontal acceleration to overcome the repulsion, while only very small drops are predicted to experience significant viscous effects. 
%
%
The results provide promising evidence and scaling laws for the mechanisms behind Leidenfrost drop repulsion, that highlight the need for further three-dimensional simulations and experimental study in order to better understand this recently reported phenomenon.
\end{abstract}

\maketitle

\section{Introduction}

The phenomenon of Leidenfrost drops was first reported in the  18\textsuperscript{th} century \cite{leidenfrost1756aquae}, where it was noted that liquid drops do not touch solid surfaces that are significantly hotter than the liquid's boiling point.
Leidenfrost drops are also seen to last significantly longer than expected, explained by an evaporative flow from the liquid lifting the drop off the plate and creating a gap between the liquid and the solid, filled with vapour. 
The long lifetime of the liquid drop is due to the layer of gas having a lower heat conductivity than the liquid
\cite{biance2003leidenfrost}, so acting as a thermal insulator and protecting the liquid from the hot plate's effects. 
%
%
Although Leidenfrost drops have been discussed for centuries, many studies have been limited to results using high-speed photography, \textit{e.g}\ \cite{burton2012geometry}, or applications of the lubrication approximation~\cite{pomeau2012leidenfrost}.
Meanwhile, recently proposed applications include small-scale pumps using waste heat \cite{wells2015sublimation}, and since the liquid phase lasts much longer (and is much more mobile) while it floats on a layer of gas \cite{vakarelski2011drag,quere2013leidenfrost},
 a ratchet-like substrate has been used to successfully move drops in desired directions (or even `up-hill') \cite{linke2006self}. 
This recently discovered method of self propulsion could find extensive applications in microfluids, and has attracted significant attention in literature \cite{lagubeau2011leidenfrost, dupeux2011viscous, cousins2012ratchet, marin2012capillary}.

Experiments involving Leidenfrost drops are often trivial to perform, since the drops evaporate slowly and are mobile whilst easy to control. A common assumption is that the drops floating on a layer of vapour are in a non-wetting situation, with an effective contact angle of $180^{\circ}$ \cite{quere2013leidenfrost}. This results in the `outer' solution for a levitating drop being comparable to that of a non-wetting sessile drop on a super-hydrophobic surface \cite{snoeijer2009maximum}. Meanwhile, non-wetting drops display an interesting property whereby they can remove themselves from a substrate at the point of coalescence \cite{boreyko2009self}. This occurs spontaneously without any external force, due to a release of surface energy. Assuming two approximately spherical drops of equal radius on a super-hydrophobic surface, and that the change in surface energy is entirely converted to kinetic energy directed away from the substrate, the theoretical vertical `take-off' velocity at coalescence has been predicted to scale as $\sqrt{\sigma/\rho r}$, where $\sigma$, $\rho$, and $r$ represent surface tension, density, and radius of the drop respectively. Data from the coalescence of Leidenfrost drops has been compared with the non-wetting theory, where a similar scaling trend for milimetric Leidenfrost drops was found as for micrometric non-wetting drops \cite{boreyko2009self}.

Droplet coalescence has become of considerable interest \cite{boreyko2009self, dietz2010visualization, miljkovic2012jumping, rykaczewski2013multimode}, since coalescors are widely used in industry: for example in water purification, oil refinery, food processing, \textit{etc}. In most applications, relying on gravity or other external forces to remove successively coalesced drops is undesired and so drop detachment via surface energy release has been proposed as an effective solution, combined with the hydrophobicity of a surface. One of the drawbacks of using flat (or rough) hydrophobic substrates is the liquid-solid contact area gives rise to a drop-substrate adhesion, making the drops stick to the surface. A recently proposed remedy is to use hydrophobic fibres \cite{zhang2015self}, where the contact area is considerably smaller, and so too the adhesive effect. Research on this topic has even extended into the nano-scale region where it has been shown that the mechanism of drop detachment should be similar, if not identical, since the liquid bridge formed at coalescence still makes contact with the solid substrate \cite{liang2015coallesence}.

Leidenfrost drops have been used in coalescence experiments to replace non-wetting drops \cite{liu2014self}, and the results compared to the numerical studies of the non-wetting case \cite{liu2014numerical}. It has been reported that the vertical momentum changes are largely down to a surface energy release combined with the inability of the resulting larger drop to permeate `into' the substrate, extending previous work on non-wetting drop self detachment \cite{boreyko2009self,sellier2011self}; and this mechanism is thought to be the same for the Leidenfrost case. 
Interestingly, despite the comparisons, the coalescence of Leidenfrost drops has not been thoroughly studied. It has been reported that Leidenfrost drops display a very strong short-range repulsion, which can prevent coalescence \cite{snezhko2008pulsating} --- remarkably even allowing drops to bounce off one another. This would signify a considerable difference to the non-wetting drops, and question the validity of equating the two systems of drop coalescence.
In the experiments performed by Liu \textit{et al}, Leidenfrost drops were made to coalesce by moving downhill towards each other, presumably acquiring sufficient momentum to overcome the short-range repulsion which was not reported \cite{liu2014self}. A comparison was then made with initially static non-wetting drops; and only vertical momentum changes were focused on. It appears there have been few, if any, non-experimental studies of Leidenfrost drop coalescence.

In light of the short-range repulsion observed \cite{snezhko2008pulsating}, Leidenfrost drop coalescence demands more detailed study. 
Numerically simulating liquid drops levitating on a cushion of evaporated vapour is non-trivial because of the system's complex and multi-scale nature,
which includes (fast) variations in temperature and several coupled dynamic processes.  
In this work, two-dimensional coalescing Leidenfrost drops are numerically simulated using the recently developed dynamic van der Waals theory (DVDWT) \cite{onuki2007dynamic}, presented to describe
one-component liquid-vapour systems with an inhomogeneous temperature field and liquid-vapour phase transition.
Using a phase-field model means that the velocity, density, and temperature fields are governed by a single set of equations, rather than requiring any pre-specified conditions to be satisfied at the interface (such a the rate of evaporation). 
The primary drawback of the phase-field method is the necessity that the interfacial region be resolved, requiring an extremely fine mesh. 
However, since the vapour layer under a Leidenfrost drop is typically extremely thin, a detailed mesh is required anyway in order to resolve the dynamics under the drop.
The DVDWT has become more attractive recently as computational ability has grown and become affordable, while it has notably been used to simulate Leidenfrost drops in previous studies; successfully resolving the dynamics in the vapour layer \cite{xu2013hydrodynamics, taylor2016numerical}. It has also been employed to
study other complex two-phase fluid problems, such as pool boiling \cite{xu2014single,taylor2016thermal}, liquid spreading \cite{teshigawara2010spreading}, droplet migration \cite{xu2012droplet}, \textit{etc}.
Simulating Leidenfrost drop coalescence using the DVDWT will allow for some remarks to be made on the vapour layer dynamics as well as the mechanism by which Leidenfrost drops rise (and fall) post-coalescence. Attention will then be turned to the repulsion between drops and the conditions required in order that the drops overcome the reported short-range force opposing coalescence. 

In order to consider the Leidenfrost effect, an appreciable vertical gravitational acceleration is required. This is different from the simulations of non-wetting drops done previously, where gravity is generally neglected since these drops are generally formed by condensation and typically extremely small.
Since previous comparisons of Leidenfrost drops are often made with \textit{initially static} non-wetting drops in close proximity, and the repulsion between Leidenfrost drops is extremely short-range; the focus will be the vapour layer dynamics of initially momentum-less Leidenfrost drops coalescing through a horizontal applied acceleration. This will allow for a broad range of relative drop sizes to be considered, although a finely-tuned simulation set-up will also be necessitated.

In the following sections, the DVDWT is briefly reviewed along with a description of the numerical method used and the simulation set-up. The 2D results for coalescing Leidenfrost drops are then presented; the overall drop coalescence process is shown followed by the drop velocity profile, and the drop height, as the systems evolve. This is followed by a study of the vapour pressure influence on the coalescence and drop jumping process. 
A study of the coalescence-resisting properties is also performed in order to identify a scaling law for the repulsion. It is found that the repulsive force is sourced from the vapour pressure of a common ``pseudo cavity'' region, that exists before the drops begin to coalesce --- in concurrence with intuition. Finally, we conclude with some remarks and comments.

%
%
%
%
\section{Modelling and Simulations}

This section begins by briefly reviewing the DVDWT \cite{onuki2005dynamic, onuki2007dynamic}: a diffuse interface model that does not need the liquid-vapour interfacial dynamics to be specified beyond the hydrodynamic equations in the bulk region
\cite{teshigawara2010spreading, laurila2012thermohydrodynamics, xu2012thermal}.
The resulting drawback is the high computational cost as the interfacial thickness ($\sim 1 \textrm{nm}$ for water) becomes
a length scale to be resolved.
The numerical method used as well as a brief overview of the boundary conditions used (a full account can be found in the references \cite{liu2012hydrodynamic}) is then presented, before the details of the simulation set-up is discussed.

%
%
%
%
\subsection{Dynamic van de Waals theory} 
A homogeneous monotomic (one-component) fluid is characterised by 
the scale of attractive interaction energy $\varepsilon$ and the molecular volume $v_0=a_{\textrm{vdw}}^3$, where $a_{\textrm{vdw}}$ is a molecular length similar to molecular diameter.
The Helmholtz free energy density is $f(n,T) = k_B T n \left[ \ln(\lambda_{th}^3 n) - \ln(1-v_0 n) -1 \right] - \varepsilon v_0 n^2$,
where $\lambda_{th}$ and $k_B$ are the thermal de Broglie
wavelength ($\lambda_{th}=\hbar\sqrt{2 \pi / m k_B T}$) and the Boltzmann constant respectively.
Here $n$ is the number density, $m$ the molecular mass, and $T$ the temperature.
The use of thermodynamic relations gives, from $f(n,T)$, the internal energy density $e$, the pressure $p$ (the equation of state),
and the entropy per molecule $s$:
\begin{align}
e &= \frac{3}{2} n k_B T - \varepsilon v_0 n ^2, \label{e}\\
p &= \frac{n k_B T}{1- v_0 n} - \varepsilon v_0 n^2, \label{P}\\
s &= -k_B \ln\left[ \frac{\lambda_{th}^3 n}{1- v_0 n} \right] +\frac{5 k_B}{2}. \label{s}
\end{align}
The critical values for $p$, $n$, and $T$ are given by $p_c=\varepsilon/27 v_0$, $n_c=1/3v_0$, and $T_c=8\varepsilon/27k_B$.  Notice that setting $v_0=0$ gives the equations for a perfect gas, as expected.

In order to describe an inhomogeneous van der Waals fluid, the system is slightly perturbed from the above description and
gradient contributions from the density inhomogeneity are introduced. Choosing the order parameter
distinguishing between liquid and vapour to be the number density, gradient contributions are introduced to the internal energy density and entropy density 
as follows:
\begin{align}
\hat{e} &= e + \frac{K(n)}{2} |\nabla n|^2, \label{gen-e}\\
\hat{S} &= n s - \frac{C(n)}{2} |\nabla n|^2. \label{gen-s}
\end{align}
Here $C$ and $K$ are positive such that the internal energy increases and the entropy decreases as a result of the inhomogeneity.
This work considers the case $C=\textrm{const.}$ and $K=0$ for simplicity; but this choice is not expected to alter the fundamental features of
the results {\cite{teshigawara2008droplet}}.
A new length scale, denoted by $l=\sqrt{C/2k_B v_0}$, arises from eqn.~\ref{gen-s}. It is of the order of magnitude of
the liquid-vapour interfacial thickness far from the critical point, and is close to $a_{\textrm{vdw}}$ \cite{teshigawara2010spreading}.

To derive the equilibrium conditions in the bulk region, the entropy in the bulk $S_b=\int \hat{S} d\mathbf{r}$ is maximised
for fixed particle number $N=\int n d\mathbf{r}$ and fixed internal energy $E_b=\int \hat{e} d\mathbf{r}$.
This leads to the equilibrium conditions: (i) the homogeneity of temperature $T$ and (ii) the homogeneity of the generalised
chemical potential $\hat{\mu}$, which is given by
\begin{align}
\hat{\mu}=\mu+\frac{M_{,n}}{2}|\nabla n|^2 - T \nabla \cdot \left( \frac{M}{T} \nabla n \right),\label{hat-mu}
\end{align}
where $M(n,T)=K(n)+C(n)T$ and $M_{,n}=\left( \partial M/\partial n \right)_T$.
The generalised pressure can be further defined through the generalised Euler equation $\hat{p}= T \hat{S} - \hat{e} + n \hat{\mu}$,
which gives
\begin{align}
\hat{p}=p-\frac{M}{2} |\nabla n|^2 +\frac{n M_{,n}}{2} |\nabla n|^2 - T n \nabla n \cdot \nabla \frac{M}{T} - M n \nabla^2 n. \label{hat-p}
\end{align}

Now the non-equilibrium hydrodynamics can be discussed in the context of assuming `local equilibrium' \cite{de2013non}.
In principle, systems that are in partial equilibria can be considered by noting that small systems reach equilibrium faster than large ones.
The whole system will reach an equilibrium state in a time-scale {$\mathcal{T}_r$}. Dividing the system into many smaller subsystems,
each subsystem is considered to reach an equilibrium state in a time-scale {$\tau_r \ll \mathcal{T}_r$}. If entropy is defined
on a time-scale $\Delta t$ with {$\mathcal{T}_r \gg \Delta t \gg \tau_r$}, then each subsystem can be considered as having reached an equilibrium state
while the system as a whole is out of equilibrium with a time-dependent entropy.

The hydrodynamic equations in the bulk region are balance equations for particle number, momentum, and energy, supplemented with
the constitutive equations (Newton's law for viscous stress and Fourier's law for heat flux).
Conservation of the number density $n$, the momentum density $\rho \vec{v}$
(with $\rho=mn$ being the mass density), and the total energy density $e_T=\hat{e}+\rho\vec{v}^2/2$ gives
\begin{align}
\frac{\partial n}{\partial t} + \nabla \cdot (\vec{v} n) &= 0, \label{mass} \\
\frac{\partial \rho \vec{v}}{\partial t} + \nabla \cdot (\vec{v} \rho \vec{v}) &=  \nabla \cdot \mathbf{M} - \rho \vec{g}, \label{momentum} \\
\frac{\partial e_T}{\partial t} + \nabla \cdot (e_T \vec{v}) &=  \nabla \cdot \left[ \mathbf{M} \cdot \vec{v} -  \vec{q}\right]-\rho \vec{g}\cdot \vec{v} ,\label{e-total}
\end{align}
with the heat flux given by $\vec{q} = - \lambda_c \nabla T$, and the total stress tensor is given by
$\mathbf{M}=-\boldsymbol{\Pi}+\boldsymbol{\sigma}$, which consists of the irreversible viscous part
$\boldsymbol{\sigma} =\eta ( \nabla \vec{v} + \nabla \vec{v}^T ) + (\zeta - 2 \eta/3) \mathbf{I} \nabla \cdot \vec{v}$ and
the reversible part denoted by $-\boldsymbol{\Pi}$.
The positive coefficients $\eta$, $\zeta$ and $\lambda_c$ denote the shear viscosity, the bulk viscosity and the heat conductivity,
respectively, while $\mathbf{I}$ is the identity. Gravitational acceleration is denoted by the vector $\vec{g}$. In the simulations, $\zeta = \eta$ is set for simplicity.
To ensure that the entropy production is free of contributions from the density inhomogeneity,
the reversible part of the stress tensor, $-\boldsymbol{\Pi}$, should take the form:
\begin{align*}
-\boldsymbol{\Pi} &= - M \nabla n \nabla n - \hat{p} \mathbf{I},
\end{align*}
in which the anisotropic part $- M \nabla n \nabla n$ leads to the liquid-vapour interfacial tension, and $\hat{p}$
in the isotropic part is defined in eqn.~\ref{hat-p}.

In order to update the temperature distribution, the entropy equation is solved (instead of the energy equation, which is reported to generate artificial parasitic flows in numerical simulations \cite{teshigawara2008droplet,onuki2007dynamic}).
Using the balance equations and standard thermodynamic relations, a balance equation
for the entropy density $\hat{S}$ is obtained:
\begin{align}
\frac{\partial \hat{S}}{\partial t} + \nabla \cdot (\hat{S} \vec{v})&
= - \nabla \cdot \vec{\mathbf{J}}^S + \frac{1}{T} \boldsymbol{\sigma} : \nabla \vec{v} - \frac{1}{T^2} \vec{q} \cdot \nabla T , \label{S-long}
\end{align}
where the total reversible entropy flux is given by
$\vec{\mathbf{J}}^S = \left[M \left( \frac{\partial n}{\partial t} + \vec{v} \cdot \nabla n \right)\nabla n+\vec{q}\right]/T$,
and the rate of entropy production is given by the last two terms in the right hand side.
To find the fluid temperature $T$ from $\hat{S}$, we combine eqn.~\ref{s} and eqn.~\ref{gen-s} to
derive a local relation between $\hat{S}$ and $T$.

%
%
%
%
\subsection{Numerical method} 

A two-dimensional rectangular system is simulated, periodic in the $x$ direction; measuring $L_x$ in the $x$ direction and $L_z$ in the $z$ direction.
The numerical scheme consists of the forward-time centred-space discretised dimensionless hydrodynamical equations~\ref{mass}, eqn.~\ref{momentum}, and eqn.~\ref{S-long},
in the two-dimensional $xz$ plane, starting from a given initial state.
The total mass is well conserved throughout the simulations, and the scheme has been found to be stable for
liquid-vapour density ratio below $\approx 5$ {\cite{xu2010contact}}. The state variables $n$, $\vec{v}$, and $T$
are defined on a non-staggered uniform Cartesian mesh, with the temperature $T$ locally determined from the entropy density $\hat{S}$.
Space discretisation is chosen to be $\Delta x=\Delta z =0.5l$, where $l$ is the liquid-vapour interfacial thickness
far from the critical point ($\approx 1{\textrm{nm}}$ for water, meaning the drops have a diameter of about a hundred nanometres only).
For the system size, $L_x=250 l$ and $L_z=210 l$ are used, sufficiently large in order to minimise the effects of the ongoing evaporation.
In the simulations, the interfacial thickness is close to $3l$, for which the spatial resolution is sufficient to resolve the associated dynamics 
--- it has been verified that finer resolutions do not have a visible impact on results \cite{teshigawara2008droplet}.

Dimensionless equations for the numerical simulations are obtained using $l=\sqrt{C/2k_B v_0}$, $V_0=\nu/l$, $\tau_0=l/V_0$, $n_0=1/v_0$ and $m n_0$
to adimensionalise the length, velocity, time, number density, and mass density respectively.
The kinematic viscosity is taken to be $\nu$; a constant independent of the local density $n$.
The shear viscosity, bulk viscosity, and heat conductivity are given by
$\eta=\nu mn$, $\zeta=\nu mn$, $\lambda_c=n k_B \nu$ respectively. Finally, the stress, entropy density, and temperature are adimensionalised using
$\varepsilon/v_0$, $k_B/v_0$, and $\varepsilon/k_B$.

The dimensionless balance equations read:
\begin{align}
& \frac{\partial n}{\partial t} + \nabla \cdot (n \vec{v}) =0, \label{continuityD}\\
& \frac{\partial \hat{S}}{\partial t} + \nabla \cdot (\hat{S} \vec{v}) = 2 \nabla \cdot (n \nabla n \nabla \cdot \vec{v})+\frac{\boldsymbol{\sigma}:\nabla \vec{v}}{T}+\left[ \nabla \cdot \left( \frac{n}{T}\nabla T \right)+\frac{n}{T^2} |\nabla T|^2 \right], \label{entropyD}\\
& \frac{\partial (n \vec{v}) }{\partial t} + \nabla \cdot (n \vec{v} \vec{v}) = \frac{1}{\mathcal{R}} \nabla \cdot \left( \boldsymbol{\sigma}-\boldsymbol{\Pi} \right) - \frac{\vec{\mathcal{G}}}{\mathcal{R}}n. \label{momentumD}
\end{align}
The two parts of the dimensionless stress tensor are given by:
\begin{align}
& \boldsymbol{\sigma} =\mathcal{R} \left[ n(\nabla \vec{v}+\nabla \vec{v}^T) + \left( \frac{n}{3} \nabla \cdot \vec{v} \right) \mathbf{I} \right], \label{sigmaD}\\
& -\boldsymbol{\Pi} =-2T \nabla n \nabla n - \hat{p} \mathbf{I}, \label{piD}
\end{align}
where the dimensionless generalised pressure is:
\begin{align}
\hat{p}= \frac{n T}{1-n}-n^2 -T |\nabla n|^2 -2Tn \nabla ^2 n.\label{pressureD}
\end{align}
Adimensionalising the hydrodynamical equations gives two dimensionless parameters.
The first one is $\mathcal{R}=m\nu^2 /\varepsilon l^2$.
To include acoustic effects for wavelengths $\mathcal{O}(l)$, a small $\mathcal{R} \ll 1$ is to be used.
Additionally, the time for sound to
travel through the system should be much longer than the viscous relaxation time $\tau_0$:
$\mathcal{R}^{1/2} L_z/l \gg 1$ with $L_z$ being the width of the system.
A common choice in literature is $\mathcal{R}=0.06$ \cite{onuki2007dynamic,teshigawara2010spreading,teshigawara2008droplet}, which is also chosen here 
(this is physically reasonable: for water $\mathcal{R} \approx 0.01$ in the liquid and $\mathcal{R} \approx 0.03$
in the vapour \cite{xu2013hydrodynamics,grigull1990tafeln}). 
The dimensionless gravitational acceleration $\vec{\mathcal{G}} \equiv m\vec{g}l/\epsilon$ is chosen to be
larger than the real value by a factor of $\approx 10^9$. This is to introduce appreciable gravitational effects
for small drops ($\sim 100 \textrm{nm}$) by decreasing the capillary length \cite{celani2009phase}. This is
necessitated by limited computational capability.
The dimensionless liquid and vapour densities at co-existence are chosen to be $n_l=0.58$ and $n_v=0.122$ respectively.

To close the above system of partial differential equations, boundary conditions are required. At the fluid-solid interface, these actually describe interfacial dissipative processes which occur
in a very thin layer near the solid. In order to avoid resolving this layer, the fluid-solid interface is modelled as a sharp interface,
with the integrated effects of the fluid-solid coupling described by the boundary conditions.
While a very general set of boundary conditions has recently been derived~\cite{liu2012hydrodynamic}, a simplified set is used here: 
(i) the no-slip boundary condition for $\vec{v}$ on the impermeable solid surfaces, 
(ii) the Dirichlet boundary condition for $T$ on the solid surface (i.e., $T=T_b$ with $T_b$ being a given constant), and 
(iii) the equilibrium condition for $n$, $\nabla_{\gamma} n = 0$ where $\nabla_{\gamma}$ is the spatial derivative normal to the solid surface. 
Note during the formation of the Leidenfrost drop (during a transient period not studied here, when the drop begins to
lift off the substrate), the full boundary conditions are expected to be crucial in any detailed study of the initial vapour layer
formation \cite{bernardin1999leidenfrost}.

%
%
%
%
\subsection{Simulation details}

\begin{figure}
\centering
\includegraphics[width= 0.8 \textwidth]{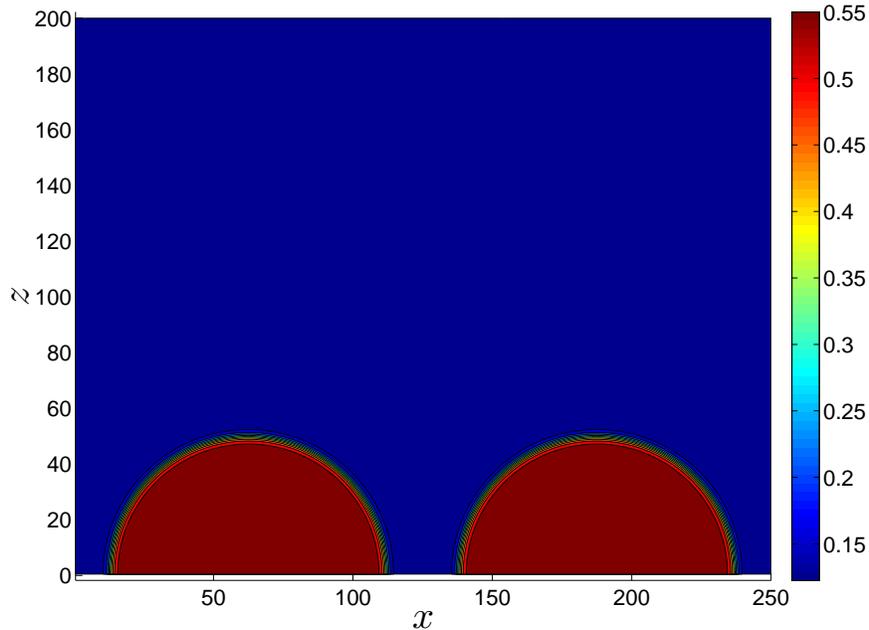}
\caption[Initial condition for the coalescence systems]{Initial condition for the system to be simulated. Two semi circular drops of liquid are allowed to reach a semi-equilibrium state in close proximity on the substrate at the co-existance temperature, without the effects of gravity. Here $x$ and $z$ are measured by $l$, and number density by $1/v_0$.}
\label{LCIC}
\end{figure}

\begin{figure}
\centering
\includegraphics[width=0.8 \textwidth]{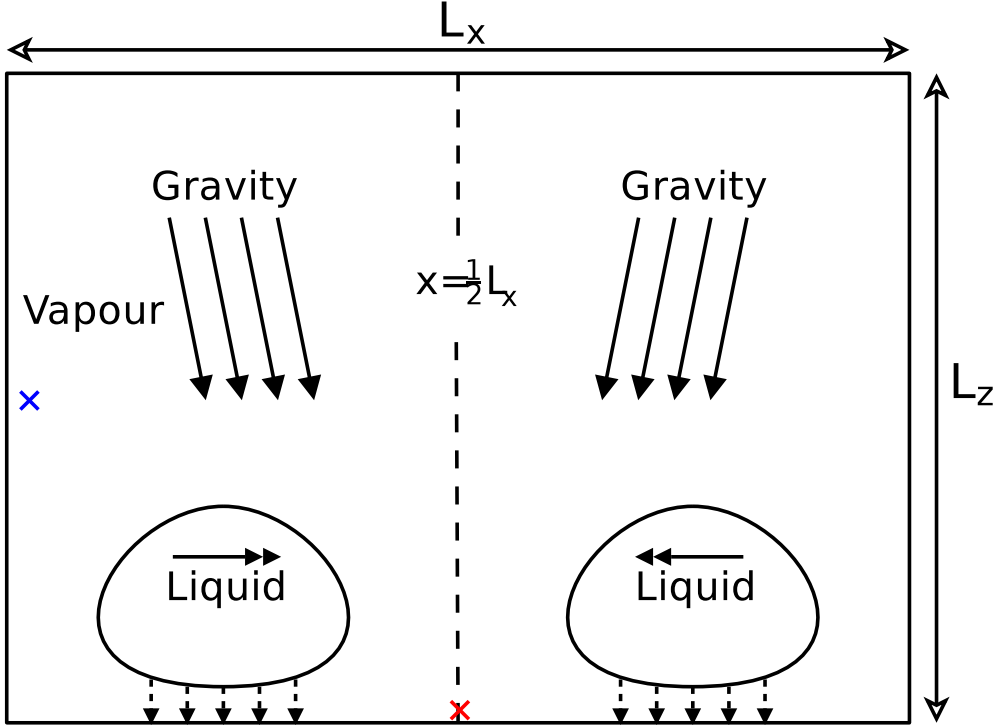}
\caption[Schematic illustration for the coalescence systems]{Schematic illustration for the system to be simulated. Notice gravity pointing toward the centre-line $x=L_x/2$. The markings on the schematic represent the various locations where pressure is measured: red and blue represent the locations where $p_S$ and $p_0$ are measured.}
\label{LCschem}
\end{figure}

In order to generate data for the initial condition, two semi circular regions of liquid of radius $50l$ are allowed to reach a semi-equilibrium state over a period of $20,000 \tau_0$ in the absence of gravity, resulting in the initial condition depicted by the system in figure~\ref{LCIC} (with $T_b=T_{cx}$). The system set-up is similar to that of a previous publication~\cite{taylor2016numerical}, where more details can be found. Here, two liquid drops are simulated subject to a varying value of the gravitational parameter $\vec{\mathcal{G}}$, which has both a vertical and horizontal component in order to cause a coalescence. The schematic of the system to be simulated is shown in figure~\ref{LCschem}, with the locations at which measurements will later be studied. The direction of the gravitational body force is drawn for reference on the diagram, it is initially given by:
\begin{align}
 \vec{\mathcal{G}}(x) = \begin{bmatrix}
        0.4\sign (x - L_x/2)\\
        1
        \end{bmatrix} \mathcal{G},
\label{GrvLC}
\end{align}
which is constant in each half of the system $x>L_x/2$ and $x<L_x/2$ pointing `inwards' towards the line $x=L_x/2$. 

Computational limitations need to be considered when choosing the body force felt by the levitating drops, set through $\vec{\mathcal{G}}(x)$. There are five important considerations: 
(i) The magnitude of the gravitational acceleration, $\mathcal{G}$ in eqn~\ref{GrvLC}, must remain nonphysically large in order for it to have an appreciable effect on the very small drops simulated here~\cite{celani2009phase}. 
(ii) The vertical acceleration, relative to the area of liquid, should ensure that the drops are of a similar nature to those simulated and experimented with previously --- that is, quasi-spherical. 
(iii) The magnitude of the horizontal component should not push the drops together too quickly, in order that the drops coalesce long after the initial stages of drop levitation are complete and any initial transient stage has elapsed~\cite{teshigawara2010spreading}. 
(iv) The time taken for the drops to meet can neither be too long, for computational resources are limited. 
(v) The drops should meet with sufficiently little momentum that the short-range repulsive force can be observed: they should be almost momentum-free immediately prior to coalescence.

In order to study different effective drop sizes, the capillary length is varied through changing the parameter $\mathcal{G}$ in eqn.~\ref{GrvLC}. Since the force is no longer normal to the substrate, this also alters the `force of attraction' between the two drops. However, this can be easily taken into account, while maintaining an otherwise identical set-up for simplicity. It is pointed out that the symmetric nature of $\vec{\mathcal{G}}$ may result in a small deformation of the larger drop post-coalescence, although this is not expected to be significant and is anticipated to have little influence on the core properties of interest.

%
%
%
%
%
%
\section{Results and Discussion}

In this section, the main results from the two-dimensional DVDWT simulations for coalescing Leidenfrost drops are presented. Initially, an overview of the timeline for the simulated system is presented, with a brief consideration of the typical drop and vapour layer dynamics immediately before drop coalescence occurs. The mass-averaged liquid drop velocities are then studied for a variety of cases, before the vapour-layer interactions between the two drops are considered and a scaling law proposed for the repulsive force.

\subsection{Overall evolution}
\begin{figure}
\centering
\vspace{-1cm}
\includegraphics[width= 0.95 \textwidth]{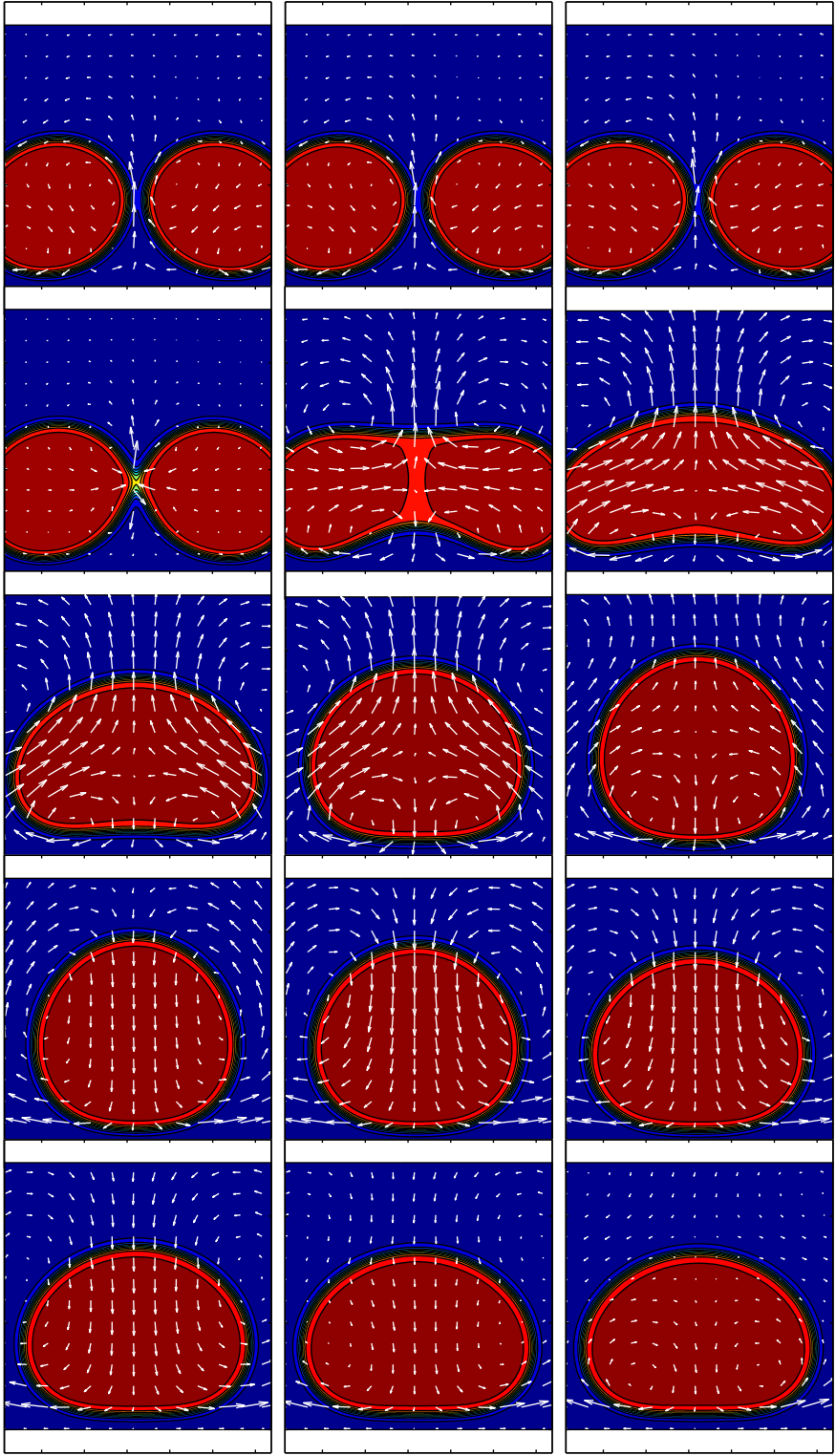}
\vspace{-2cm}
\caption[The Leidenfrost drop coalescence process]{The Leidenfrost drop coalescence process. Results for $\mathcal{G}=1 \times 10^{-5}$ are shown for the central area of the domain at $250 \tau_0$ intervals. Colour represents density while white arrows the local fluid velocity. The first frame is at IC+$4,000 \tau_0$ in the top left. The time is measured by $\tau_0 \equiv l^2/\nu$} \label{evol-LC}
\end{figure}

In figure~\ref{evol-LC}, the coalescence of two Leidenfrost drops is shown for the case of $\mathcal{G}=1 \times 10^{-5}$, 
at time intervals of $250 \tau_0$. Both the density distribution (in colour) and the velocity field (white arrows) are shown. The first frame is captured $4,000 \tau_0$ after the initial conditions are generated and gravity is turned on, and the substrate temperature increased. Only the central region $62l \le x \le 187l$, $0 \le z \le 125l$ is shown to focus on the drop-drop interactions. In the top-left panel note the non-circular shape of the drops. 
The drop deformation actually becomes stronger as the drops get closer together --- some of the evaporative vapour from the drop undersides flows through the gap between the drops in a vertical direction causing a viscous shear to the liquid, resulting in the deformation seen pre-coalescence. Note that the low viscosity ratio (in these simulations $\approx 5$) exaggerates this effect; and that a non negligible flow can be seen inside the liquid drops pre-coalescence. 
Left, second from top, the initial stages of liquid bridge formation are seen, at the point of coalescence. 
Right, second from top, the bridge is seen to have anti-symmetrically expanded: the vapour layer pressure appears to resist the downward liquid movement as the bridge thickens.

\begin{figure}
\centering
\mbox{
\includegraphics[width= 0.5\textwidth]{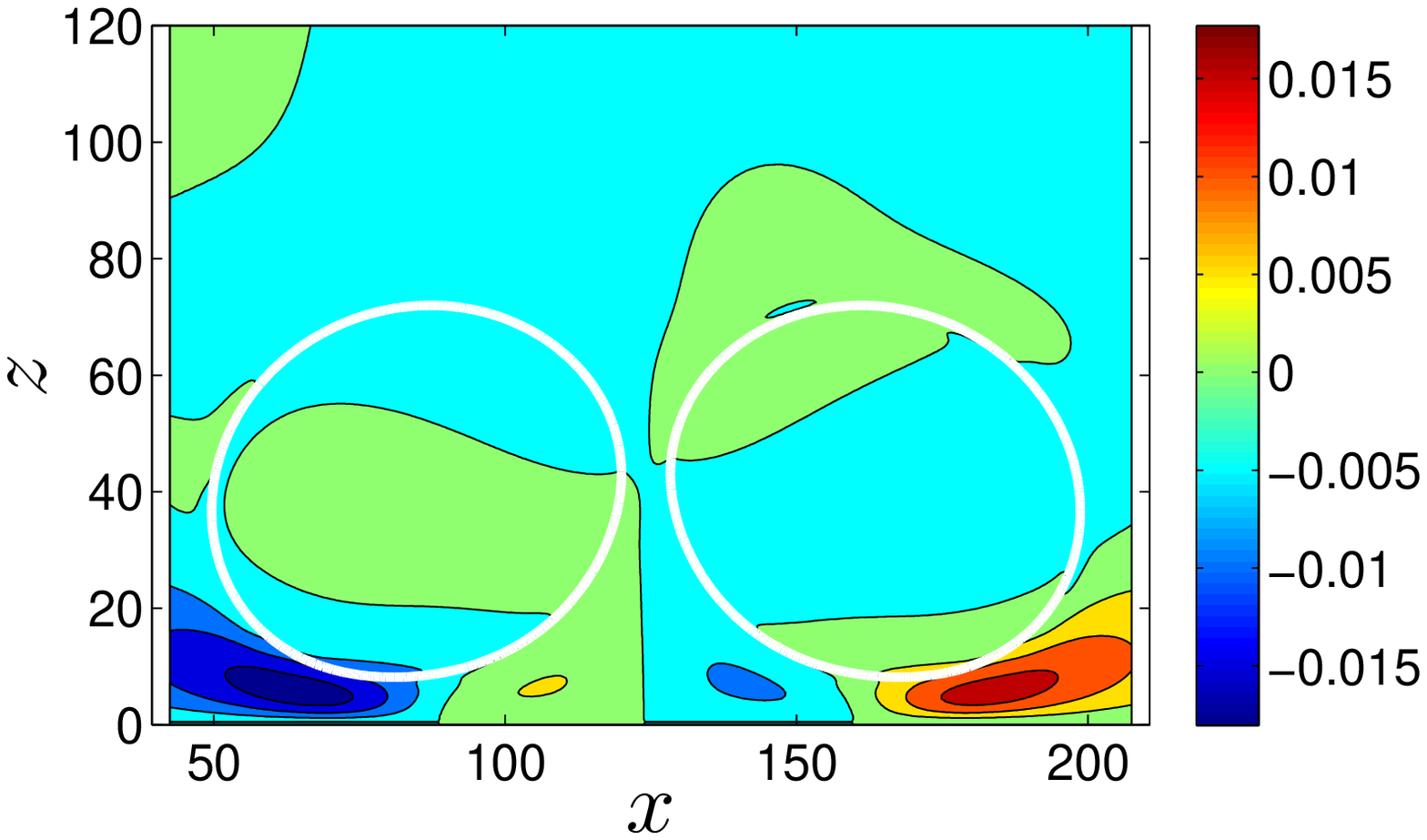}\quad
\includegraphics[width= 0.5\textwidth]{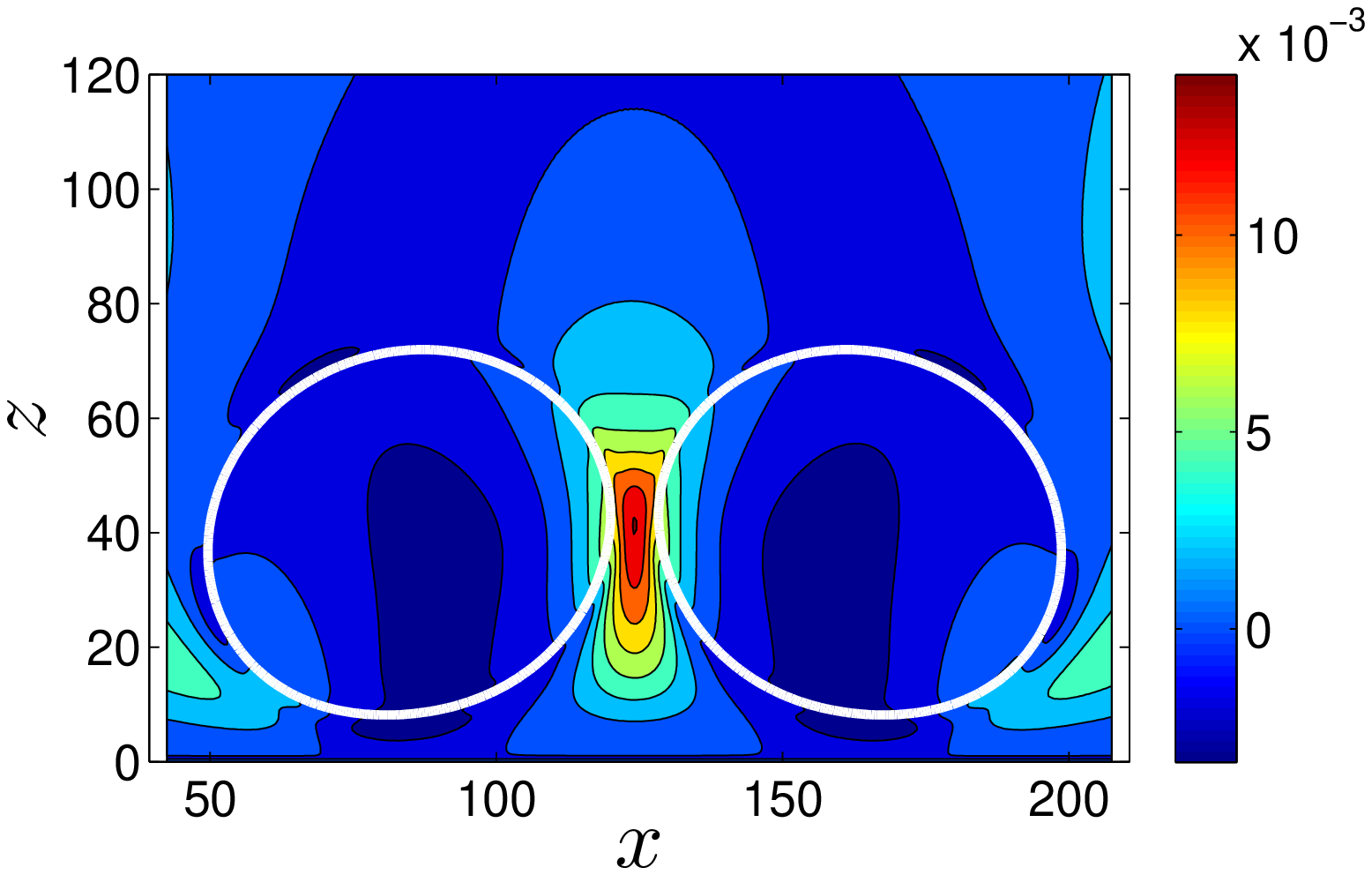}
}
\caption[Vapour Velocity]{Fluid Velocity. The $u$ (left) and $w$ (right) components of the velocity field of the fluid for the semi-steady state observed as Leidenfrost drops await coalescence. Here $x$ \& $z$ are measured by $l$, and $u$ \& $w$ by $\nu/l$.} \label{Fluid-V}
\end{figure}

The velocity of the fluid is shown in both the horizontal (left) and vertical (right) directions in figure~\ref{Fluid-V}, at time $t=3,000 \tau_0$ showing the approximately steady-state maintained until coalescence. The density contour at $n=n_i \equiv (n_l+n_v)/2$ is also shown in white. It is clear that the vapour layer dynamics are altered substantially from the case of a single Leidenfrost drop previously studied~\cite{taylor2016numerical}; here the symmetry of the vapour layer under an individual drop (particularly the $u$-component of velocity) is broken. It is seen that there is a strong vapour flow leaving the \textit{combined} cavity region at either side, but a much smaller contribution towards the central area between drops. 
This is explained by the interaction of the two vapour layers --- there is a build up of vapour in the combined cavity region, with only a small chimney between drops whereby there is an escape. This leads to an increased pressure in the entire central cavity region, which forces the vapour flow through the thin gap between drops as seen in the $w$ component of figure~\ref{Fluid-V}. 
%

\subsection{Drop velocity}

Physically, the Leidenfrost effect is a vertical force balance between liquid weight and pressure sourced from an evaporative flow, necessitating a different set up to previous studies of coalescing drops --- generally performed without any significant vertical body force \cite{sellier2011self, liu2014numerical}. After the drops coalesce and the liquid rises up due to the change in total surface energy, the coalesced drop will immediately decelerate due to gravity and then return to the substrate, and is expected to bounce slightly on the vapour layer. 
The Leidenfrost drop coalescence considered in this paper will therefore result in an overall rise in the average liquid distance from the substrate --- that is, the gravitational potential of the liquid will be increased to a maximum before decreasing again. Since the Leidenfrost drops are always detached from the substrate, the term `jump' in this context refers to any initial rise in liquid that is later eventually reversed by gravity.

\begin{figure}
\centering
\includegraphics[width= \textwidth]{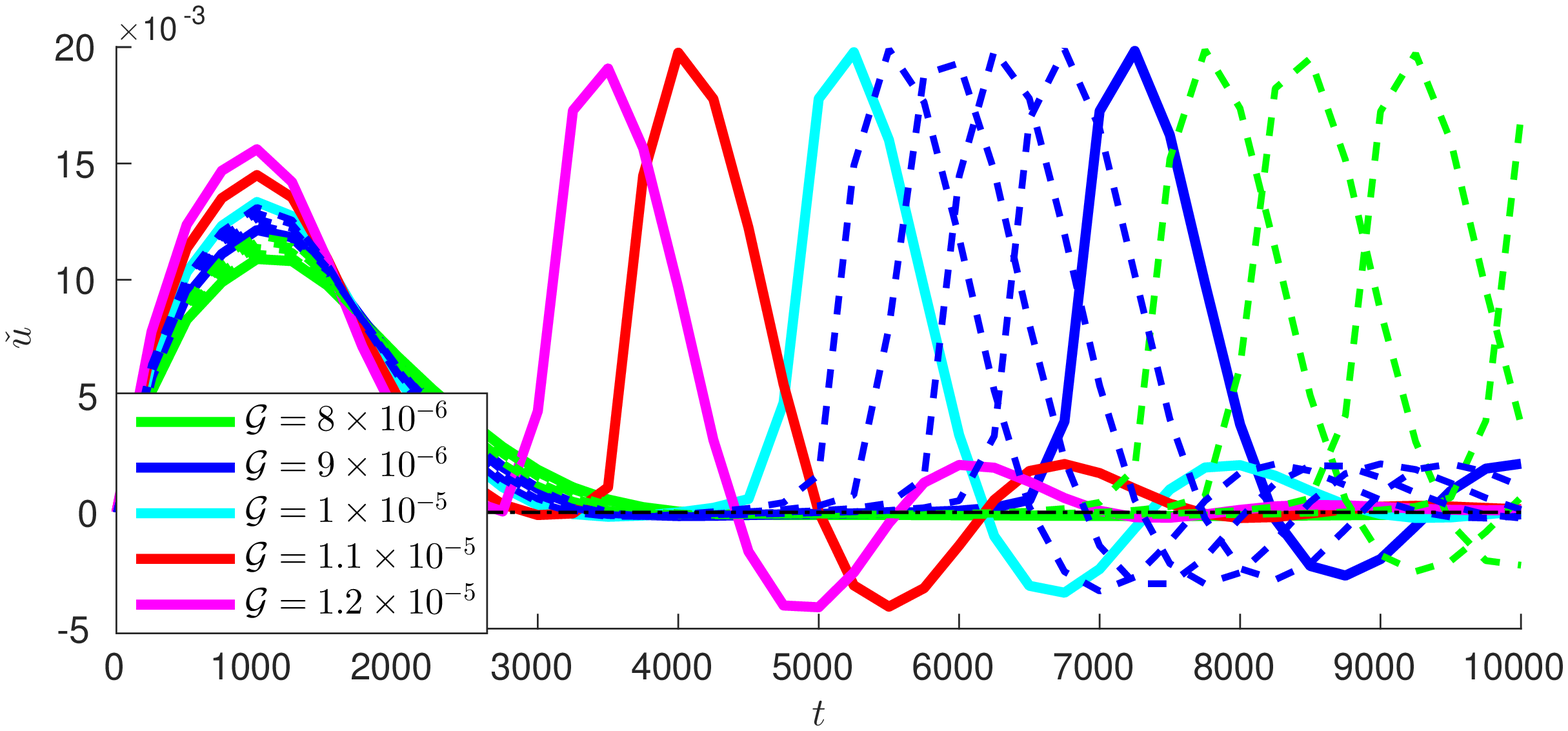}\\
\includegraphics[width= \textwidth]{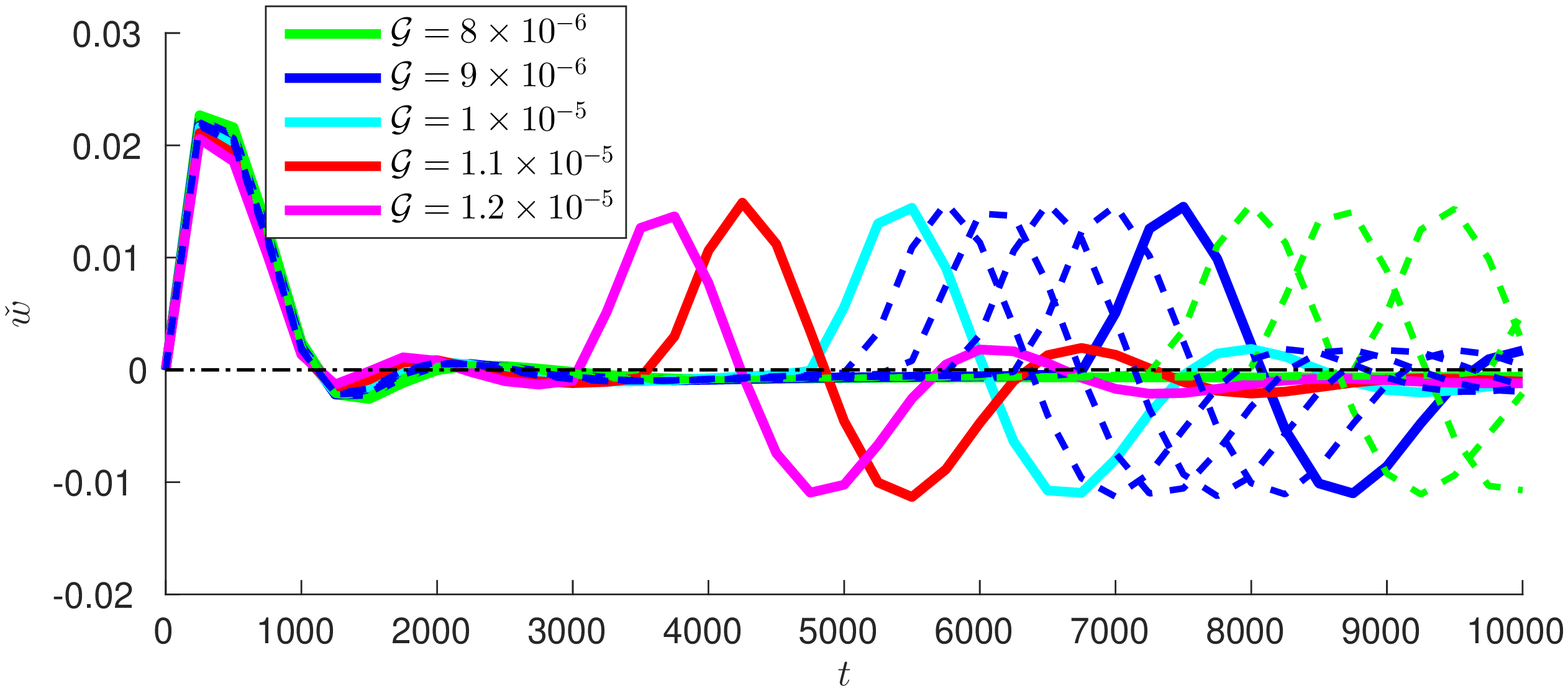}
\caption[Drop Velocity]{Mass averaged drop velocities $\check{u}$ (upper) and $\check{w}$ (lower) as drops merge, for various values of $\mathcal{G}$. Dashed lines indicate values of $\mathcal{G}$ between those of the solid lines, at equal increments of $2 \times 10^{-7}$. The velocity is measured by $\nu/l$, and the time by $\tau_0 \equiv l^2/\nu$.} \label{Drop-V}
\end{figure}

The 2D mass-averaged drop velocities, $\check{u}$ and $\check{w}$, are now considered. Here, the 2D mass-averaged equivalent of a scalar quantity $\psi$ is defined as $\check{\psi}$:
\begin{align*}
\check{\psi} = \frac{\iint_{\textrm{liquid}} n \psi dA}{\iint_{\textrm{liquid}} n dA} = \frac{\iint_{n>n_i} n \psi dA}{N_l},
\end{align*}
where $N_l$ is the number of particles in the 2D drop. 
The mass-averaged tangential and vertical velocities, $\check{u}$ and $\check{w}$, are plotted for the left-hand drop, for a number of gravitational parameters, in figure~\ref{Drop-V}. As expected, the early stages of the simulation consist of the vapour layer forming, and the drop leaving the substrate, stabilising quickly in the vertical direction during the initial $1,000 \tau_0$. Meanwhile, the applied horizontal body force accelerates the drops towards each other. However, the drop-drop repulsion quickly retards them as they approach one another, and before coalescence they are almost stationary (both $\check{u}$ and $\check{w}$ are close to zero). This is a requirement imposed on the choice of $\vec{\mathcal{G}}$ since residual momentum would make studying any very short-range forces potentially challenging.

Coalescence occurs at a later time after varying amounts of evaporation have taken place, resulting in a sharp increase in both $\check{u}$ and $\check{w}$. Viscous dissipation is fast in the simulations so there is little oscillation and strong damping. 
Upon coalescence the new larger liquid mass quickly changes shape, beginning to take the form of a large Leidenfrost drop as seen in figure~\ref{evol-LC}, \textit{i.e} a cavity space under the coalesced drop is formed. As the coalescence process evolves, a more circular shape is seen in figure~\ref{evol-LC} as the surface energy is minimised. Finally, after falling due to gravity, a quasi-circular shape is achieved and the underside is deformed, as expected for Leidenfrost drops, with the gap between substrate and liquid similar to that pre-coalescence. The changes in drop shape can be easily matched to the changes in average drop velocity, for example the rise and fall in $\check{w}$ here is a quantification of the change in vertical speed of the liquid seen in figure~\ref{evol-LC}.

\begin{figure}
\centering
\includegraphics[width= \textwidth]{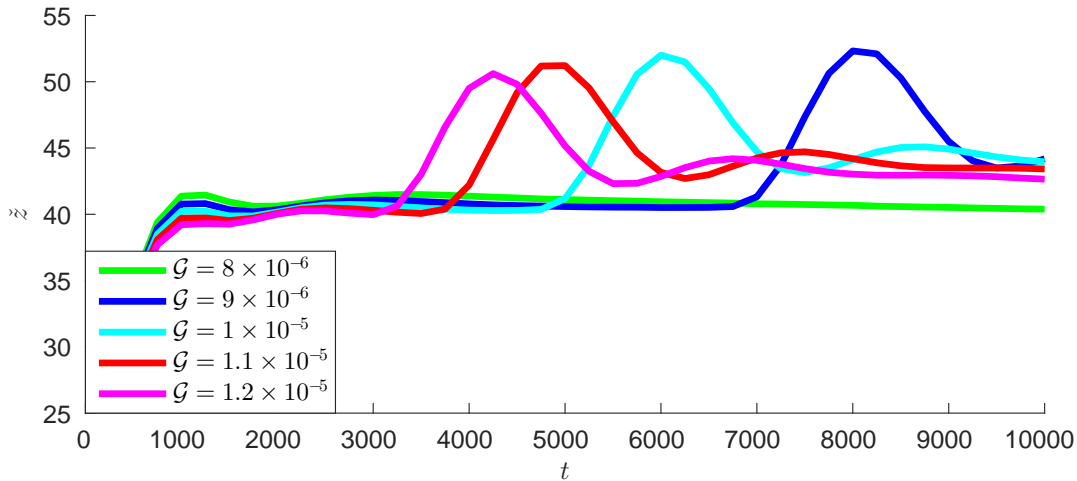}
\caption[Coalescence height evolution]{Drop height evolution. The evolution of the height of the centre of gravity, $\check{z}$, of the liquid drop. The length is measured by $l$, and the time by $\tau_0 \equiv l^2/\nu$} \label{DropZ-LC}
\end{figure}

In figure~\ref{DropZ-LC} the mass-averaged drop height, $\check{z}$, is plotted as a function of time through the coalescence process for several choices of $\mathcal{G}$. Since the initial increase in $\check{z}$ is later partially reversed (due to gravity forcing the liquid back toward to substrate, although $\check{z}$ remains above its initial level since larger Leidenfrost drops have higher centres of mass), it is confirmed that Leidenfrost drops still experience a `jump' resulting from coalescence as in the non-wetting case. After the initial `jump', gravity along with the vapour layer pressure and viscous damping act together to stabilise the drop height.
Note that decreasing values of $\mathcal{G}$ lead to an increasing maximum value seen for $\check{z}$. This is primarily due to the lower vertical gravitational acceleration, and is consistent with a release of surface energy causing the increase in $\check{z}$, as with the cases of non-wetting drops.

\subsection{Vapour layer interactions}

Here, the vapour layer pressure prevents the liquid phase from lowering and touching the substrate --- contrasting with the ``liquid bridge impacting substrate'' mechanism underlying the jump of coalescing non-wetting drops \cite{boreyko2009self}. 
For Leidenfrost drops, the vapour layer pressure is expected to contribute to the system dynamics in two ways: (i) From figure~\ref{Fluid-V}, it is predicted the short-range repulsion of Leidenfrost drops is controlled by the vapour pressure which also leads to the vapour flow through the gap between drops. (ii) As the bridge is formed it moves closer to the substrate, as seen in figure~\ref{evol-LC}, and hence the shape of the coalesced drop will remain coupled with the pressure similar to the case of a single Leidenfrost drop. It is expected the maximum size of a coalesced drop (before a chimney-like instability splits the liquid phase) will be below that of a single steady Leidenfrost drop; since the vapour pressure is temporarily above the steady-state value during coalescence.

\begin{figure}
\centering
\includegraphics[width= \textwidth]{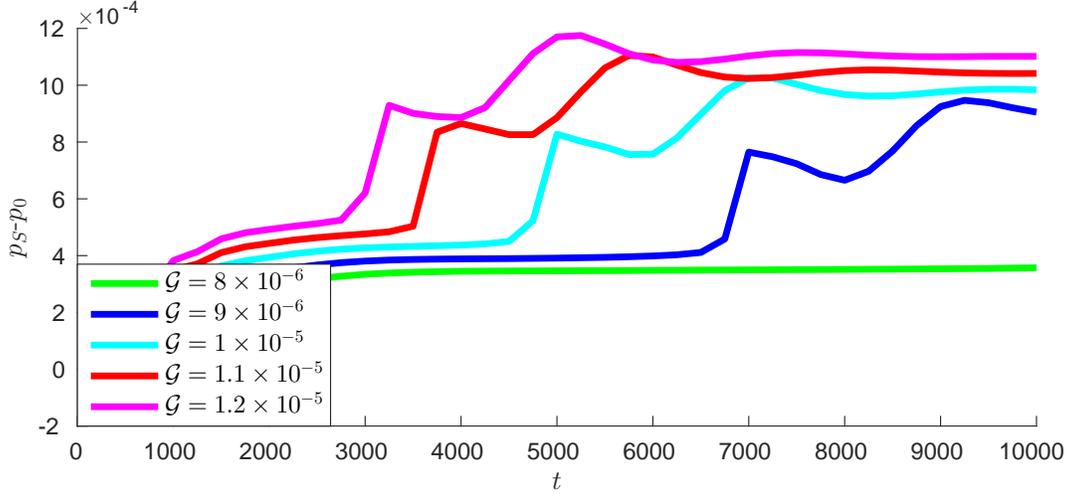}
\caption[Vapour layer pressure changes]{Vapour layer pressure changes. The pressure difference between the centre of the common vapour layer (at $x=L_x/2$), $p_S$, and the ambient pressure, $p_0$, is plotted through time as coalescence occurs. The locations for pressure measurements are shown on figure~\ref{LCschem}. The pressure is measured by $\epsilon/v_0$, and the time by $\tau_0 \equiv l^2/\nu$} \label{P-Diffs}
\end{figure}

Figure~\ref{P-Diffs} shows the evolution of the pressure difference $p_s-p_0$ (the positions of which are shown in figure~\ref{LCschem}). Note that $p_s$ is measured at $x=L_x/2$ --- which is only under the liquid phase \textit{after} coalescence has begun. 
Before coalescence, while the drops are almost stationary, it can be seen the pressure in the combined cavity is almost constant, and significantly different to the ambient pressure, $p_0$. 
The pressure under the merging drops is then seen to change substantially during coalescence. As the drops merge and the liquid bridge expands, the space under the now-single liquid mass shrinks rapidly, immediately causing a sharp initial increase in vapour pressure at $p_s$. 
Subsequently, the vapour layer pressure is seen to oscillate before reaching a final steady value under the drop. 
The oscillations are coupled with the liquid `bouncing' on the vapour layer as seen in figure~\ref{DropZ-LC}. 
After the final stages of the drop coalescence (once a new nearly-steady state is reached), the pressure profile in the vapour layer is as a single, larger, Leidenfrost drop: it peaks around the centre of the vapour layer, at $x=L_x$.
The overall (vertical) momentum changes of the liquid drop cannot be attributed to the pressure changes in the vapour layer: the liquid momentum changes are found to be much more substantial than the pressure-sourced force. This infers that the vapour layer does not play a significant role in the liquid `jump' during coalescence besides preventing liquid-solid contact as the bridge expands.
%
%
However, the feedback mechanism where evaporation gets stronger as the interface (rather than $\check{z}$) approaches the substrate still exists and does contribute to maintaining the levitating nature of the drop, deforming it's underside in an active manner different to the non-wetting case. 
It is deduced that while the vapour layer pressure changes in order to prevent liquid-solid contact, the overall momentum changes are caused by the minimisation of surface energy. The mechanism behind the \textit{jump} in liquid mass is therefore similar to that of non-wetting drops, since the vapour pressure does not contribute significantly to the total liquid acceleration. This confirms the theory proposed in the literature \cite{boreyko2009self}: Leidenfrost drops are propelled upwards upon coalescence by the same mechanism (and scaling laws) as non-wetting drops, \textit{i.e} a release of surface energy.

Now turning to the short-range repulsion between drops; it is found that there is insignificant opposing evaporation in the thin gap between drops --- contradicting the mechanism previously suggested~\cite{snezhko2008pulsating}. Therefore, the repulsion is taken to be due to the build up of vapour in the combined space beneath both drops. This was established in the pressure-driven vapour flow through the gap between drops, seen in figure~\ref{Fluid-V}~(right). 
Moreover, prior to coalescence the pressure in the common space between drops, $p_s$, is seen to remain almost constant in figure~\ref{P-Diffs} (\textit{e.g} the green curve), despite the drops continually shrinking due to evaporation. This pressure is expected to act on an increasingly small length of interface as the drops get smaller, reducing the force opposing coalescence until it becomes insufficient to overcome the applied horizontal body force, which changes at a much slower rate.

In Pomeau's application of the lubrication approximation, the scaling law for the pressure in the vapour layer under a vanilla Leidenfrost drop is found to be $R^{5/2} \rho_l g R_l^{-3/2}$~\cite{pomeau2012leidenfrost}. 
The pressure difference between each side of a 2D drop in close proximity with another is expected to adhere to a similar dependence, since the gap between drops is very small. Using the expression for $R_l$: 
$$R_l=\left( \frac{\eta \lambda \Delta T}{g L \rho_l \rho_v} \right)^{\frac{1}{3}},$$ 
which represents the drop radius magnitude at which the lubrication approximation breaks down~\cite{celestini2012take}, the vapour pressure contribution from the drops to the combined cavity region is proposed to scale as $\Delta p_L$:
\begin{align*}
\Delta p_L = g_z^{3/2} R^{5/2} \sqrt{\frac{L \rho_l \rho_v}{\eta \lambda \Delta T}}.
\end{align*}
Meanwhile, the force pushing a single drop toward the other is denoted $F_x =M_D g_x$, where $M_D$ is the (2D equivalent) mass of the drop and $(g_x,g_z)$ the (horizontal, vertical) acceleration applied. Therefore, the first time when the following forces become of equal magnitude (as $R$ decreases through evaporation) represents the maximum radius for the net force to act inwards;
\begin{align*}
F_x &= \pi R^2 \rho_l g_x, \\
F_p \approx \pi R \Delta p_L &\sim g_z^{3/2} R^{7/2} \pi \sqrt{\frac{L \rho_l \rho_v}{\eta \lambda \Delta T}},
\end{align*}
where $F_p$ is the force due to the vapour pressure in the combined cavity beneath two drops, acting outwards on the inside quarter of the drops interface. It has also been assumed that the drops are small enough they can be characterised by their equivalent radius where $R=(N_l/n_l \pi)^{1/2}$, which is reasonable in the simulations presented here. 
Notice that since only two-dimensional simulations are performed, that do not display axis symmetry, it is two-dimensional ``equivalent forces'' that are being compared; \textit{i.e} Newtons/metre. This will impact the relative strength of the repulsive force, making it much more significant in 2D than it may be in 3D.

For initially momentum-less drops to begin to join, the tangential forces $F_x$ \& $F_p$ must be balanced, meaning $R$ and $\vec{g}=(g_x,g_z)$ should satisfy:
\begin{align}
R^2 \rho_l g_x &\sim g_z^{3/2} R^{7/2} \sqrt{\frac{L \rho_l \rho_v}{\eta \lambda \Delta T}}, \label{fullR-Balance}
\end{align}
or in terms of $R_l$;
\begin{align}
R^2 \rho_l g_x &\sim R_l^{-\frac{3}{2}} g_z \rho_l R^{\frac{7}{2}}. \label{Rl-Balance}
\end{align}
Since $R_l/R$ is typically very small, the repulsive force can be considered very strong; most Leidenfrost experiments are performed for $R \gg R_l$.

In the numerical simulations, the dimensionless parameter $\mathcal{G} \equiv mgl/\epsilon$ is used to denote the gravitational acceleration, as well as measuring $R$ in terms of $l$. By rescaling the parameters $R$ and $\vec{g}$ in eqn.~\ref{fullR-Balance} using the scalings in the DVDWT simulations, an estimate can be made for the constants in the balance:
\begin{align}
\underbrace{\frac{\epsilon l \rho_l}{m}}_{A} \bar{R}^2 \mathcal{G}_x \sim  \underbrace{\sqrt{\frac{L \rho_l \rho_v \epsilon^3 l^4}{\eta \lambda \Delta T m^3}}}_{B} \mathcal{G}_z^{3/2} \bar{R}^{7/2}. \label{ABFBal}
\end{align}
Now $\mathcal{G}_{x}$ and $\mathcal{G}_{z}$ are the dimensionless parameters used in the simulations (for the horizontal and vertical component of $\vec{\mathcal{G}}$ respectively) and $\bar{R}$ is measured in terms of $l$. Table~\ref{Material-Values} gives some physical values (in SI units) for water and nitrogen at co-existence, where $D_{Tv}$ is the thermal diffusivity of the vapour phase, in order for an order-of-magnitude estimate to be made for the slope $A/B$ in eqn.~\ref{ABFBal}.

\begin{table}
\centering
\begin{tabular}{c||c|c||c}
\hline
\hline
Substance & Water & Nitrogen & Average\\
\hline
$T_{cx}$ & $603$ & $115$ & $360$\\
$\eta_v$ & $22 \times 10^{-6}$ & $9.3 \times 10^{-6}$ & $16 \times 10^{-6}$\\
$D_{Tv}$ & $1.3 \times 10^{-6}$ & $9 \times 10^{-7}$ & $1.1 \times 10^{-6}$\\
$\lambda_v$ & $95 \times 10^{-3}$ & $21 \times 10^{-3}$ & $58 \times 10^{-3}$\\
$L$ & $1.1 \times 10^{6}$ & $1.2 \times 10^{5}$ & $6.1 \times 10^{5}$\\
$m$ & $3.0 \times 10^{-23}$ & $2.3 \times 10^{-23}$ & $2.7 \times 10^{-23}$\\
$\rho_l$ & $52$ & $490$ & $270$\\
\hline
\hline
\end{tabular}
\caption[Material properties of water \& nitrogen]{Table of material properties for water and nitrogen at co-existence \cite{grigull1990tafeln, sychev1987thermodynamic}. 
\label{Material-Values}}
\end{table}

Taking average values for water and nitrogen at the co-existence temperature, estimations for $A$ and $B$ can be found (using liquid-vapour density ratio $n_l/n_v = 4.8$, vapour values for $\eta$ \& $\lambda$, and $\Delta T=0.11 T_{cx}$) noting that $\mathcal{R} \equiv m \nu^2/\epsilon l^2 = 0.06 $ was chosen and $\nu \approx 7 \times 10 ^{-7}$ (leading to the Prandtl number being $\approx 5$). Using these values it is found that the gradient between vapour layer repulsion and tangential body force is predicted to be $\approx 6$.

\begin{figure}
\centering
\includegraphics[width= \textwidth]{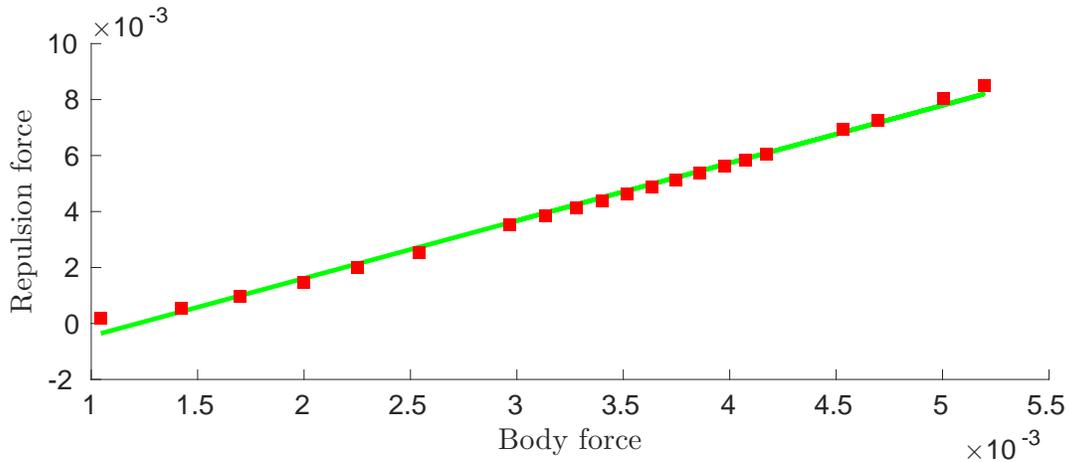}
\caption[Leidenfrost force balance]{Minimum radius for net drop repulsion. Red squares show data points at the time of coalescence; when the attractive and repulsive forces approximately balance. The solid green line shows a linear best fit, where the intercept and gradient are $-2.5 \times 10^{-3}$ and $2.1$ respectively. In the simulations, $R_l \approx 0.2 \mathcal{G}^{1/3}$ corresponding to a body force of between $\approx 3.1 \times 10^{-4}$ and $\approx 3.9 \times 10^{-4}$; smaller than those points shown here.} \label{vapour-body}
\end{figure}

Figure~\ref{vapour-body} shows the force balance in eqn.~\ref{ABFBal} at the point of coalescence from over twenty different simulations (where each simulation represents a different choice of $\vec{\mathcal{G}}$), where the solid green line has a gradient of $2.1$; which is fitted to the data. 
The fitted gradient is in good agreement with the theoretical value predicted, and certainly within an acceptable level of error since the material values can only be very crudely estimated from co-existence properties.
In order to get the complete range of data shown in figure~\ref{vapour-body} there has been a variety of magnitudes of $|\vec{\mathcal{G}}|$ used and different ratios between the horizontal and vertical components. The initial ratio of $2:5$ in eqn~\ref{GrvLC} was selected to allow a good range of magnitudes to result in coalescence in the time frame considered for the drop size chosen, but this ratio does not allow smaller drops to coalesce from a momentum-less semi-steady state. It becomes technically demanding to simulate large drops (large $\mathcal{G}_z$) for the following two reasons: (i) It is more challenging to ensure the drops are outside a transient period with almost zero momentum yet in very close proximity. (ii) The ratio between tangential and horizontal gravity becomes a very sensitive factor for larger values of $|\vec{\mathcal{G}}|$.

Notice that for very small drops, towards the left-hand side of figure~\ref{vapour-body}, there is a noticeable deviation from the linear fit. In fact if these small drops are excluded, the linear fit is much improved. Moreover, a constant offset has been included in plotting the fitted green line, despite eqn.~\ref{ABFBal} predicting a zero intercept. 
Since the simulations here represent very small drops (with a very small viscosity ratio), the effects of viscous resistance may not be negligible. As the drops evaporate the drop radius will recede, making the gap between drops wider (in the absence of any competing body forces).  
Since the rate of the radius change becomes larger as the drops become smaller, the linear fit taking only vapour pressure and body forces into account cannot be made to pass through the origin; the smallest drops will need to overcome an increasingly significant viscous drag originating from moving at the speed of the receding interface,~$\approx \dot{R}$, in order that the gap between drops does not widen.

By considering a Stokes-like drag force in the horizontal $x$-direction, an estimate on the size of drops for which viscosity becomes significant can be made. 
In order for the gap between drops to not widen, the drops must move horizontally at a speed greater than that of the receding radius. The effective viscous `force' acting on a 2D circular solid disk is therefore taken to linearly depend on $\eta \dot{R}$ \cite{proudman1957expansions, kaplun1957asymptotic}. This assumes a high viscosity ratio, leading to the expectation that the viscous forces will be far greater in the simulations than those predicted (since the simulated viscosity ratio is only $\approx 5$). To approximate $\dot{R}$, the total rate of evaporation is equated with the change in circular mass as;
\begin{align}
\dot{M} \equiv 2 \pi \rho_l R \dot{R} \sim (R^3 g_z \rho_l \rho_v)^{\frac{1}{4}} \frac{\pi}{\eta} \left( \frac{\eta \lambda \Delta T}{L} \right)^{\frac{3}{4}}, \label{Mass-shrink}
\end{align}
using the scaling law for the mass of fluid leaving the vapour layer under a Leidenfrost drop~\cite{pomeau2012leidenfrost}.
Utilising eqn.~\ref{Mass-shrink} the viscous drag term is found to be negligible in real life experiments compared to vapour pressure repulsion, up until $R \sim R_l$ at which point the Leidenfrost regime changes significantly and the scaling laws here would no longer apply.  In the simulations, $R_l \approx 0.2 \mathcal{G}^{1/3}$ corresponding to a body force of $\approx 4 \times 10^{-4}$, which is smaller than the smallest drop simulations shown in figure~\ref{vapour-body}.

\section{Concluding Remarks}

The dynamic van der Waals theory has been used to simulate two-dimensional liquid-vapour systems containing a pair of Leidenfrost drops under the influence of a directed body force. It was found that upon coalescence, the small drops do not touch the substrate and remain supported by the vapour layer pressure. The previous prediction that the change in vertical momentum for Leidenfrost drops is similar to that of non-wetting drops (\textit{i.e} a release of surface energy) was confirmed, and the mechanism whereby the coalesced drop rises up was found to be due to factors beyond the vapour pressure, which peaks during the coalescence process. 

The results confirmed experimental observations that Leidenfrost drops experience a remarkably strong short-range repulsion. It was seen that the force resisting coalescence scales the same way as the pressure in the vapour layer under a single drop, with good agreement for the two-dimensional numerical cases. Where the body force pushing the two liquid drops together is small enough, they initially do not coalesce. Instead they come to rest in close proximity, evaporating until they become sufficiently small that the length of interface exposed to the vapour layer (and therefore vapour pressure) results in an insufficient force to continue to dominate the applied horizontal acceleration. In the simulations performed, there was a low viscosity ratio and a very small length scale used, resulting in viscous effects becoming apparent for the smallest drops. An approximation for the force balance including viscous effects showed that viscous drag only becomes significant for extremely small drops (of the order of $R \sim R_l$), and is unlikely to influence the drop joining process in macroscopic experiments.

Although the simulation results are insightful, they are only applicable to cases where the drops are initially stationary and in a semi-steady state awaiting coalescence. In the simulated systems this results from the initial drops being too large to coalesce, so they approach one another, where the repulsion allows them to remain close until they have evaporated to become small enough to join. Immediately pre-coalescence, there is negligible overall drop momentum. 
In previous Leidenfrost coalescence experimental studies, the drops are not momentum-less \cite{liu2014self}. They are generally rolled towards each other; meeting with enough momentum to easily overcome the repulsive force. The length scale on which the repulsive force becomes significant has not been studied here, however this is a crucial future work to consider since it would allow for an estimate to be made for the momentum required in order to allow coalescence. 
In addition, fluctuations in substrate topology, drop shape, or contaminants in the fluid (which all exist in experiments) are not modelled here. These small perturbations may result in significant changes in the interface profile and local momentum changes could then trigger the initial stages of coalescence.

Crucially, the simulations presented here consider only two-dimensional drops and it is not possible to extend the results to higher dimensionality. This is significant since in three dimensions, although the pressure in the vapour layer is expected to scale similarly, the the curvature of the drop interface will result in the gap between drops varying with both $y$ as well as $z$. As a result, the 2D scaling law for the force balance can not be translated into higher dimensional space. Full three-dimensional simulations are needed in order to develop this work further, however the mechanism behind Leidenfrost drop repulsion and Leidenfrost jumping upon coalescence are expected to be the same.

\section*{Acknowledgements}
The author would like to thank Tiezheng Qian for helpful guidance. This work has been made financially possible by the Hong Kong RGC.
\bibliographystyle{apsrev4-1}

\end{document}